\documentclass[12pt]{article}
\usepackage{graphics,epsfig}
\begin{document}
\begin{center}
{\bf ON THE PROPERTIES OF THE  $^{229}$Th ISOTOPE}

\vspace{1.0cm}
V. I. Isakov\footnote{E-mail Visakov@thd.pnpi.spb.ru}

\vspace{0.3cm}
{\it Petersburg Nuclear Physics Institute, Gatchina 188300, Russia,}

\vspace{0.3cm}
National Research Centre  Kurchatov Institute

\vspace{1.0 cm}
{\it \bf A b s t r a c t}
\end{center}

{\small Electromagnetic properties of the deformed neutron-odd nucleus $^{229}$Th are investigated in the
framework of the unified model, with primary emphasis upon the properties of the low-lying isomeric state.}

\vspace{1.0cm}
On the basis of detailed analysis of $\gamma$-transitions in $^{229}$Th attendant $\alpha$-decay of
$^{233}$U, it was established the existence in the daughter nuclei $^{229}$Th of the low-lying level
with the excitation energy of only a few eV. This is the most low-lying state known by now. The next
one is the level 1/2$^+$ in $^{235}$U, with the  excitation energy equal to 76.5 eV. The latest
experimental data \cite{Beck07} point to the value of the excitation energy equal to $\sim$ 7.6 eV.
In the paper \cite{Lars16}, the authors detected conversion electrons arising from the decay of
this level. In this way, they proved that this level really exists, and it`s energy is above the threshold
of ionization of neutral atom of Th, which is equal to $\sim$ 6.3 eV. Together, the half-life of this level
equal to $7(\pm 1)\,\, \mu$s was measured in \cite{Seif17}.  However, the energy of this state is
not yet measured in the direct experiment.

Here, we carry out theoretical analysis of the characteristics
of $^{229}$Th, and make an attempt to describe decay properties of it`s low-lying levels, as well as
to propose an alternative way for excitation of the above-mentioned state in the reaction of the
Coulomb excitation.

In Fig.1  we show experimental scheme of levels and the decay scheme for the low-lying states in
$^{229}$Th, that are known by now from the experiment \cite{Gold89,nds08}. Here, one can easily observe
rotational bands characteristic to the deformed nuclei. Thus, we perform theoretical analysis for this
deformed neutron-odd nuclei basing on the ideas of the unified model proposed in the papers \cite{Bohr52}--
\cite{Bohr55}.

\begin{figure}[h]
\centering
\includegraphics[width=0.5\hsize]{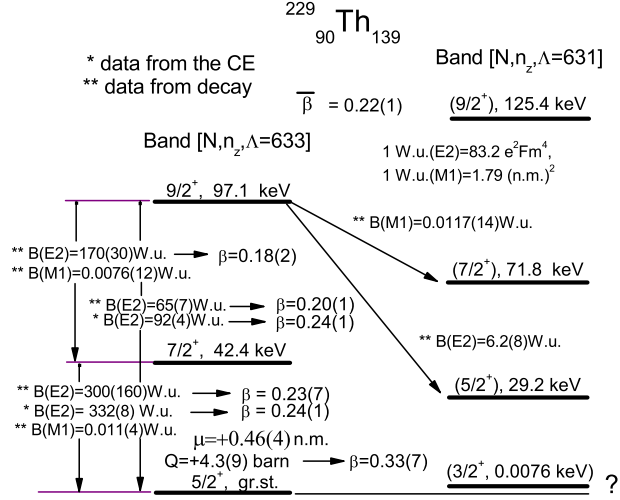}
\caption{\small
Low-lying levels in  $^{229}$Th}
\end{figure}

The wave function of the axially-symmetric odd nuclei in the framework of the unified model reads as

\begin{equation} 
\Psi^J_{MK}\ =\ \sqrt{ \frac{2J+1}{16\pi^2}} \Big[D^J_{MK}(\theta_{i})
\cdot\chi_K  + D^J_{M-K}(\theta_{i})\cdot \overline{\chi^J_K}\Big]\,.
\end{equation}

Second term in (1) provides symmetry of the wave function to  reflection relatively the plane
orthogonal to the symmetry axis, while

\begin{equation} 
\chi_K = \sum_{N\ell\Lambda s} x_K(N\ell\Lambda s) |N\ell\Lambda s\rangle\,,\quad
\overline{\chi^J_K} = \sum_{N\ell\Lambda s} (-1)^{J-\ell-1/2} x_K (N\ell\Lambda s)|
 N\ell-\Lambda-s\rangle\,.
 \end{equation}

In (1) and (2) $\chi_K$ are Nilsson  orbitals \cite{Nilsson55} that represent the decomposition of the
single-particle functions of the axially-symmetric deformed potential over the spherical-symmetric functions,
$\Lambda$ and $s$ are projections of orbital moment and spin on the symmetry axis, $K = \Lambda + s$.

We define  reduced transition matrix elements and  reduced transition rates by the relations
\begin{eqnarray} 
&& \hspace*{-0.5cm}
\langle J_2M_2|\hat m(\lambda\mu)| J_1M_1\rangle\ =\ (-1)^{J_2-M_2}
\left( \begin{array}{ccc} J_2 & \lambda & J_1\\ -M_2 & \mu & M_1 \end{array} \right)
\langle J_2\|\hat m(\lambda)\|J_1\rangle\,,
\nonumber \\
&& \hspace{2.0cm} \langle J_2\|\hat m(\lambda)\|J_1 \rangle = (-1)^{J_2-J_1}\langle J_1\|\hat m(\lambda)\|J_2\rangle\,.
\end{eqnarray}

\begin{equation}
B(\lambda;J_1 \to J_2) = \frac{\langle J_2\|\hat m(\lambda)\|J_1\rangle^2}{2J_1+1}\,,\,\, B(\lambda;J_1 \to J_2)=
\frac{2J_2+1}{2J_1+1} B(\lambda; J_2 \to J_1)\,.
\end{equation}

For $E2$ transitions we have
\begin{equation}
\hat{m}(E2,core)^2_{\mu}=D^2_{\mu 0}(\theta_{i})\cdot\frac{3}{4\pi} ZR^2\cdot\beta|e| , \ {\rm while}
\  \,\,\hat{m}(E2,s.p.)^2_{\mu} = \sum_{\nu} D^2_{\mu \nu}(\theta_{i})\cdot\hat{m}(E2,intr.)^2_{\nu}\,.
\end{equation}

Then,  we obtain
\begin{eqnarray} 
&& \hspace*{-0.5cm}
\langle \Psi^{J_2}_{K_2}\, \| \hat m(E2,core) +\hat m(E2,s.p.)
\|\Psi^{J_1}_{K_1} \rangle\ =
\nonumber \\
&& = \quad (-1)^{J_2-J_1} \sqrt{(2J_1+1)}\ C^{J_2K_2}_{20\,J_1K_1}
 \delta(K_1,K_2)\delta(\alpha_1,\alpha_2)\,\frac{3}{4\pi}\, |e| \cdot ZR^2\cdot \beta\ +
 \nonumber\\
&& +\quad  (-1)^{J_2-J_1}\ \sqrt{\frac{5(2J_1+1)}{4\pi} }\cdot\!\!
\sum_{N\ell\Lambda (1,2)} x_{K_2}(N_2\ell_2\Lambda_2s_2)\,x_{K_1}
(N_1\ell_1\Lambda_1s_1)\,.(u_1u_2 - v_1v_2)\ \times
  \nonumber\\
&& \times \quad \sqrt{\frac{2\ell_1+1}{2\ell_2+1}}C^{\ell_20}_{20\ell_10}\Big[\delta(s_1,s_2) C^{J_2K_2}_{2(K_2-K_1)J_1K_1}
C^{\ell_2\Lambda_2}_{2(K_2-K_1)\ell_1\Lambda_1} \ +
\nonumber\\
&&+\quad  \delta(s_1,-s_2)(-1)^{J_2-\ell_2-1/2}C^{J_2-K_2}_{2(-K_2-K_1)J_1K_1} C^{\ell_2-\Lambda_2}_{2(-K_2-K_1)\ell_1\Lambda_1}\Big]
\,e_{\rm eff}\, \langle2|r^2|1\rangle\,.
 \end{eqnarray}

In (6) $u$ and $v$ are the coefficients of the Bogoliubov transformation, that accounts the superfluid correlations,
while $e_{\rm eff}$ is the effective quadrupole charge for the odd particle.

Quadrupole moment of state is  expressed via the reduced $E2$ matrix elements by the relation
\begin{equation}
 Q_2(J)\ =\ \sqrt{ \frac{16\pi\,J(2J-1)}{5(J+1)(2J+1)(2J+3)}}\ \langle\,J\|\hat m (E2)\| J\,\rangle\,.
\end{equation}

Consider now  $M1$ transitions. Because of the particle-hole polarization arising from the spin-dependent
interactions between the nucleons, ``bare'' values of gyromagnetic ratios in nuclei renormalize. In addition,
this polarization leads to the appearance of the additional tensor term in the single-particle $M1$ operator
which ``opens''  $l$-forbidden transitions in spherical nuclei. In this way, the $M1$ transition operator
in our case reads as

\begin{equation} 
\hat m(M1)^1_\mu\ =\ \sqrt{\frac3{4\pi}}\,\mu_N\, \Big[g_R\hat J+(g_\ell-g_R) \hat\ell
+(g_s-g_R)\hat s +\delta\hat \mu(M1,tens.)\Big]^1_\mu\,,
\end{equation}

\begin{equation}
{\rm where} \ \,\delta\hat \mu(M1,tens.)^1_\mu\ =\ \kappa r^2[Y_2 \otimes\hat s]^1_\mu \cdot \tau_3\,.
\end{equation}

In (9) $\tau_3=+1$ for neutrons $(n)$ and $\tau_3=-1$ for protons $(p)$; $\kappa=-0.031$\,\,fm$^{-2}$; $g_\ell(p)\approx1.1$,
$g_\ell(n)\approx0.0,$ \ $g_s(p)=3.79$, $g_s(n)=-2.04$, $g_R=Z/A=90/229=0.393.$
The values of parameters $g_l, \ g_s$ and $\kappa$ were defined by us before \cite{Art82,Isa16} from the description of
magnetic moments as well as $l$-allowed and $l$-forbidden $M1$ transition rates in spherical nuclei, both near
and far from the closed shells. As a result, we obtain the formula for the reduced $M1$ transition matrix element:

\begin{eqnarray}
&& \hspace*{-0.5cm}
\langle \Psi^{J_2}_{K_2}\|\hat m(M1,core) +\hat m(M1,s.p.)\|\Psi^{J_1}_{K_1}\rangle\ =
\nonumber\\
&&= \quad g_R\,\,\mu_N\,\delta(K_1,K_2)\delta(J_1,J_2)\delta(\alpha_1,\alpha_2) \sqrt{\frac{3J_1(J_1+1)(2J_1+1)}{4\pi}}\ +
\nonumber\\
&& +\ (-1)^{J_2-J_1} \sqrt{\frac{3(2J_1+1)}{4\pi}}\,\mu_N\sum_{N\ell\Lambda s(1,2)}x_{K_2}(N_2\ell_2\Lambda_2s_2)
x_{K_1}(N_1\ell_1\Lambda_1s_1)\,.(u_1u_2+v_1v_2)\times
\nonumber\\
&&\times\Bigg\{ (g_\ell-g_R)\sqrt{\ell_1(\ell_1+1)}\,
\delta(n_1,n_2)\delta(\ell_1,\ell_2) \bigg[\delta(s_1,s_2) C^{J_2K_2}_{1(K_2-K_1)J_1K_1}\ \times
\nonumber\\
&& \times\ C^{\ell_2\Lambda_2}_{1(K_2-K_1)\ell_1\Lambda_1}
+  \delta(s_1-s_2)(-1)^{J_2-\ell_2-1/2} C^{J_2-K_2}_{1(-K_2-K_2)J_1K_1}
C^{\ell_2-\Lambda_2}_{1(-K_2-K_1)\ell_1\Lambda_1}\bigg]\ +
\nonumber\\
&&+\ (g_s-g_R)\frac{\sqrt3}2\,\delta(n_1,n_2)\delta(\ell_1\ell_2)\bigg[\delta(\Lambda_1,\Lambda_2)
C^{J_2K_2}_{1(K_2-K_1)J_1K_1} C^{1/2\,s_2}_{1(K_2-K_1)1/2\,s_1}\ +
\nonumber\\
&& +\quad \delta(\Lambda_1,-\Lambda_2)(-1)^{J_2-\ell_2-1/2} C^{J_2-K_2}_{1(-K_2-K_1)J_1K_1}
C^{1/2-s_2}_{1(-K_2-K_1)1/2\,s_1}\bigg]\ -
\\
&&-\ \kappa\langle2|r^2|1\rangle \bigg[C^{J_2K_2}_{1(K_2-K_1)J_1K_1}
\langle\ell_2\Lambda_21/2\,s_2|[Y_2\otimes\hat s]^1_{(K_2-K_1)}\,|\ell_1\Lambda_11/2\,s_1\rangle\ +
\nonumber\\
&&+(-1)^{J_2-\ell_2-1/2} C^{J_2-K_2}_{1(-K_2-K_1)J_1K_1} \langle\ell_2-\Lambda_21/2-s_2|
[Y_2\otimes\hat s]^1_{(-K_2-K_1)} |\ell_1\Lambda_11/2\,s_1 \rangle\bigg]\Bigg\} .
\nonumber
\end{eqnarray}

Here, $K_1=\Lambda_1+s_1$, $K_2=\Lambda_2+s_2$\,, while

\begin{eqnarray} 
&& \hspace*{-0.5cm}
\langle\,\ell_2\Lambda_21/2\,s_2|[Y_2\otimes\hat s]^1_\mu\,|\ell_1\Lambda_1 1/2\,s_1]\,\rangle\ =\
\frac32 \sqrt{\frac{5(2\ell_1+1)}{2\pi}}\, C^{\ell_20}_{\ell_10\,2\,0}\ \times
\nonumber\\
&& \hspace*{-1.0cm}\times \sum_{j_1j_2} \sqrt{(2j_1+1)} C^{j_2K_2}_{j_1K_11\mu} C^{j_2K_2}_{\ell_2\Lambda_21/2s_2}
C^{j_1K_1}_{\ell_1\Lambda_11/2s_1}
\left\{ \begin{array}{ccc} \ell_2 & 1/2 & j_2\\  \ell_1 & 1/2 & j_1\\  2 & 1 & 1 \end{array}
  \right\},\ \mu=\Lambda_2+s_2-\Lambda_1-s_1;
\end{eqnarray}

Magnetic moments of states are defined by the relation

\begin{equation}
  \mu_J\ =\ \sqrt{\frac{4\pi J}{3(J+1)(2J+1)}}\ \langle\,J \|\hat m(M1)\|J\,\rangle\,.
\end{equation}

For the $E2$ transitions between the states of the same rotational band, we may in formula (6) take
into account only collective part of the matrix element, as the single-particle one gives only
a small contribution. Then, we have standard formulas for the quadrupole moments of states and for the
transition rates \cite{Bohr53,Bohr55}, where the result depends only on the deformation parameter
$\beta$ and the entering values of $J$ and $K$:

\begin{equation}
Q_2(J,K) = \frac{3K^2-J(J+1)}{(J+1)(2J+3)}\,Q_0;\,\,   
     Q_2(K=J) = \frac{J(2J-1)}{(J+1)(2J+3)}\,Q_0\, ;\,
     Q_0\ = \frac{3}{\sqrt{5\pi}}|e|ZR^2\cdot \beta.
\end{equation}

\begin{equation} 
B(E2;J+1,K\to J,K)\ =\ \frac{3K^2(J+1+K)(J+1-K)}{J(J+1)(J+2)(2J+3)} \cdot \frac5{16\pi}\,Q^2_0\, ,
\end{equation}

\begin{equation}
B(E2;J+2,K\to J,K)\ =\ \frac{3(J+2+K)(J+1+K)(J+2-K)(J+1-K)}{(2J+2)(2J+3)(J+2)(2J+5)}\cdot \frac5{16\pi}\,Q^2_0\,.
\end{equation}

By using  experimental data shown in Fig.1 and formulas (13)--(15), one can easily define the magnitude of the
deformation parameter $\beta$ which average value turns out to be $\overline\beta \approx 0.22$. This is close to
the magnitude of $\beta$, that corresponds to maximal value of the binding energy $B$ in $^{229}$Th obtained in calculations
\cite{Gogny}, which were performed in the  Hartree--Fock--Bogoliubov approach with the Gogny interaction. This
value of $\beta$ was used by us in our calculations that involve the ``intrinsic'' function $\chi$.

Consider now  transitions between the states of different bands $|(J_1,J_1^{'}) K_1\rangle \to |(J_2,J_2^{`})K_2\rangle$,
where the initial as well as final states have different values of $(J,J^{`})$, but the same values of $K$.   We see
from formula (6)  that in case of the $E2$-transitions matrix element contains the multiple $(u_1u_2-v_1v_2)$, which
value is very sensitive to small variation of the single-particle scheme, especially  when the
entering single-particle orbitals are close to the Fermi level. This is just the case under consideration.
In addition, the value of the effective quadrupole charge $e_{\rm eff}$ is rather indefinite here, as it is not
clear, what part of the quadrupole transition strength should be included  in the single-particle mode after
taking into account rotation of the core in the obvious way. Thus,  direct calculations of the $E2$ transition
matrix elements are not trustworthy here. However, one can easily see
from formulas (6) and (10), that if the multipolarity of radiation $\lambda$ satisfies the condition
$K_1+K_2 > \lambda$, as it takes place if we consider $E2$ and $M1$ trasitions between the bands $[633]$ and $[631]$,
then we have the relation \cite{Alaga55}

\begin{equation} 
B(\lambda; J'_1K_1 \to J'_2K_2) = \frac{[C^{J'_2K_2}_{\lambda(K_2-K_1)J'_1K_1}]^2}{[C^{J_2K_2}_{\lambda(K_2-K_1)J_1K_1}]^2}
B(\lambda; J_1K_1 \to J_2K_2).
\end{equation}

As we know from the experiment the value of $B(E2;J_1=9/2,K_1=5/2 \to J_2=5/2,K_2=3/2)$ = 6.2(8) W.u., we can define in this
way all interband $E2$-transition matrix elements.

The situation is different in case of  $M1$ transitions. Here, both collective and single-particle parts of the $M1$
transition matrix element (10) give comparable contributions even in cases of transitions within the same
rotational band. In this case,  multiple $(u_1u_2+v_1v_2)$ is close to unity, while
the values of $g_s$ and $\kappa$ are known. Thus, calculations of $M1$ transition matrix elements were performed in the
obvious way, both for  interband transitions and for transitions within the same band.

Results of our calculations of the $E2$ and $M1$ electromagnetic characteristics of the $^{229}$Th  are shown in Tables 1--3.
As one can see from Table 1, the magnitude $B(M1;9/2,5/2\to7/2,3/2)$ obtained in our calculations is about four times less
than the average experimental value shown in \cite{nds08}. Note however that different experimental values of this quantity
differ from each other tenfold more than experimental errors.  One can obtain in our calculations average value \cite{nds08}
only if we use $g_{s}(n)\approx g_{s}(n)_{\rm free}$, which contradicts generally established conception. Note that if we
borrow the value of the $M1$ single-particle transition matrix element
$\langle\chi_{5/2}|\hat m(M1,s.p.)|\chi_{5/2}\rangle$ from the experimental data on the $|9/2,5/2\rangle
\to |7/2,5/2\rangle$ and $|7/2,5/2\rangle \to |5/2,5/2\rangle$ $M1$ transitions, we in the best cases (by taking the proper
sign of the matrix element) have the worse agreement with the experiment on the value of the ground-state magnetic moment
of $^{229}$Th, as compared to results of direct calculations.

By using data on transition rates shown in Table 1, $B(M1;3/2,3/2\to5/2,5/2)=0.0108 \,\mu_N^2$ and
$B(E2;3/2,3/2\to5/2,5/2)=8.0$ W.u., and  the values of the conversion coefficients
for the 0.0076 keV $\gamma$-transition \cite{Mila07} (also the private communication of M.B. Trzhaskovskaya),
we find the half-lives for this transition equal to $T_{1/2}(M1) =  5.9\cdot\,10^{-6}\,$s and
$T_{1/2}(E2) = 2.7\cdot\,10^{-3}\,$s (including electron conversion). Here, conversion coefficients
are very large: $\alpha^{M1}_{tot}(0.0076\,\,{\rm keV})\approx 1.4\cdot\,10^9$ and
$\alpha^{E2}_{tot}(0.0076\,\,{\rm keV})\approx 1.2\cdot\,10^{16}$. It is important that at such small
transition energies, conversion coefficients rapidly grow with decrease of the transition energy (approximately,
$\alpha^{M1}_{tot} \sim 1/(\Delta E)^{3-\epsilon}$ and $\alpha^{E2}_{tot} \sim 1/(\Delta E)^{5-\epsilon}$, where $\epsilon \sim 0.05$).
{\em As a result, the half-life of the state of interest at such small transition energies in practice does not depend on
energy, but only on the transition matrix element}.
In \cite{Striz91,Tkalya15} one can find other evaluations  of the magnitude of $T_{1/2}(3/2,3/2\to5/2,5/2)$
\footnote{The latest theoretical estimations for transition rates in $^{229}$Th  are in  \cite{Minkov}}.

\vspace{0.5cm}
Below, we discuss  the problem of population of  the above-mentioned isomeric state by the method different from
$\alpha$ and $\beta$-decays. In the paper \cite{Jeet15},  authors proposed the method which employs  synchrotron radiation,
while in \cite{Inamura05} the authors suggested pumping $^{229m}$Th by the hollow-cathode discharge.
Here, we consider the chance for excitation of the isomeric state in the Coulomb excitation, the process that was
proposed for the first time in \cite{Mart52} and elaborated in details in \cite{Alder561}.

\begin{table}[h]
\caption{Reduced transition rates for the interband $E2$ and $M1$ transitions between the low-lying levels
of the bands $[N_1n_z(1)\Lambda_1]=[633]$ and $[N_2n_z(2)\Lambda_2]=[631]$ in $^{229}$Th. Calculations
of the $E2$ transition rates, shown in the W.u., were based on formula (16), where the
experimental value of $B(E2;J_1=9/2\,K_1=5/2\, [633]\to J_2=5/2\,K_2=3/2\, [631])=6.2(8)$
was used as the normalization factor.
Here, W.u.($E2$) =83.2 $e^2$fm$^4$. Results on the $B(M1)$ values are  shown in the units of $\mu_N^2$
(1 W.u.$(M1)=1.79\,\, \mu_{N}^2$),
and they were obtained by calculations basing on formula (10) with
$\beta = 0.2,\, g_s(n, \rm eff) = -2,04$ and $\kappa = -0.031$ fm$^{-2}$. }
\begin{center}
{\small
\begin{tabular}{||c|c|c|c||c|c|c|c||}
\hline\hline

$E,M(\lambda)$ & $J_1\,K_1$ & $J_2\,K_2$ & $B(1\to2)$  &
$E,M(\lambda)$ & $J_1\,K_1$ & $J_2\,K_2$ & $B(1\to2)$ \\
\hline\hline

$E2$ & 9/2\,3/2 & 5/2\,5/2 & 0.53  & $E2$ & 3/2\,3/2 & 5/2\,5/2 & 8.0  \\

$E2$ & 9/2\,3/2 &7/2\,5/2 &  4.6  & $M1$ & 9/2\,3/2 & 7/2\,5/2 & 0.00072\\

$E2$ & 9/2\,3/2 & 9/2\,5/2 &  3.9 & $M1$ & 9/2\,3/2 & 9/2\,5/2 & 0.00460 \\

$E2$ & 9/2\,5/2 & 5/2\,3/2 &  6.2 [6.2(8)] & $M1$ &  9/2\,5/2 & 7/2\,3/2 & 0.00506 [0.0209(25)]\\

$E2$ & 9/2\,5/2 & 7/2\,3/2 & 0.11 & $M1$ & 7/2\,3/2 & 5/2\,5/2 & 0.00039  \\

$E2$ & 7/2\,3/2 & 5/2\,5/2 & 3.3  & $M1$ & 7/2\,3/2 & 7/2\,5/2 & 0.00413 \\

$E2$ & 7/2\,3/2 & 7/2\,5/2 & 5.7 & $M1$ & 7/2\,5/2 & 5/2\,3/2 & 0.00581 \\

$E2$ & 7/2\,5/2 & 3/2\,3/2 & 5.3  & $M1$ & 5/2\,3/2 & 5/2\,5/2 & 0.00310 \\

$E2$ & 7/2\,5/2 & 5/2\,3/2 & 0.22 & $M1$ & 3/2\,3/2 & 5/2\,5/2 & 0.01080 \\

$E2$ & 5/2\,3/2 & 5/2\,5/2 & 8.0  &      &          &          &         \\
\hline\hline
\end{tabular}
 }\end{center}
\end{table}

\begin{table}[h]
\caption{Reduced $E2$ and $M1$ transition rates between the levels inside  the $K=5/2$ and $K=3/2$ bands. Here,
the   $B(E2)$ values are in the Weisskopf units and were calculated by using  $\overline\beta
= 0.22$. Numbers in square brackets show experimental results \cite{Gold89,nds08}.
The M1  rates are in the units of $\mu^2_N$,
and they were calculated by using $g_s(n, \rm eff) =-2.04$ and $\kappa =-0.031\rm\,fm^{-2}$. }
\begin{center}
{\small
\begin{tabular}{||c|c|c|c||c|c|c|c||}
\hline\hline
$E,M(\lambda)$ & $J_1\,K_1$ & $J_2\,K_2$ & $B(1\to2)$ &
$E,M(\lambda)$ & $J_1\,K_1$ & $J_2\,K_2$ & $B(1\to2)$\\  \hline\hline

  $E2$ & 9/2 3/2 & 5/2 3/2 &  167 & $E2$ & 5/2 3/2 & 3/2 3/2 & 267 \\

  $E2$ & 9/2 3/2 & 7/2 3/2 &  109 & $M1$ & 9/2 3/2 & 7/2 3/2 & 0.0583 \\

  $E2$ & 9/2 5/2 & 5/2 5/2 &   78 [85(4)] & $M1$ & 9/2 5/2 & 7/2 5/2 & 0.0386 [0.0136(21)]\\

  $E2$ & 9/2 5/2 & 7/2 5/2 &  236 [170(30)] & $M1$ & 7/2 3/2 & 5/2 3/2 & 0.0521 \\

  $E2$ & 7/2 3/2 & 3/2 3/2 &  111 & $M1$ & 7/2 5/2 & 5/2 5/2 &  0.0266 [0.0197(72)]\\

  $E2$ & 7/2 3/2 & 5/2 3/2 &  167 & $M1$ & 5/2 3/2 & 3/2 3/2 &  0.0389 \\

  $E2$ & 7/2 5/2 & 5/2 5/2 &  279 [330(8)] &      &         &         &        \\
  \hline\hline
 \end{tabular} }
 \end{center}
  \end{table}

\begin{table}[h]
\caption {Electric quadrupole and magnetic dipole moments of the lowest states of $^{229}$Th.
Here, by calculation of quadrupole moments we used averaged value of $\overline\beta$ =0.22, while by
calculation of magnetic moments we used $\beta = 0.2,\, g_s(n, \rm eff) =-2.04$ and
$\kappa =- 0.031\rm\,fm^{-2}$.}
\begin{center}
{\small
\begin{tabular}{||c|c|c||c|c|c||}
\hline\hline
Quantity$(J\,,K)$ & Exp. & Calc. & Quantity$(J\,,K)$ &  Exp. & Calc. \\ \hline\hline

$Q_2(5/2\,,5/2)$ &  +4.3(9) barn & +2.9 barn & $Q_2(3/2\,,3/2)$ & --   &  +1.6 barn  \\

$\mu(5/2\,,5/2)$ & +0.46(4) $\mu_N$  & $+0.47 \,\mu_N$ & $\mu(3/2\,,3/2)$ &   --   &  $+0.12\, \mu_N$\\
\hline\hline
\end{tabular}}
\end{center}
\end{table}


\begin{table}[h]
\caption{Comparison between the cross sections $\sigma$ and the ``effective'' cross sections \newline
$\sigma_{\rm eff}$ for the Coulomb excitation of the $^{229}$Th levels by protons and $\alpha$-particles.}
\vspace{0.5cm}
\begin{tabular}{||c|c||c|c||c|c||c|c||}
\hline\hline
  & Energy & \multicolumn{2}{c||}{Protons, 6 MeV} & \multicolumn{2}{c||}{Protons, 10 MeV} & \multicolumn{2}{c||}{ $^4$He, 10 MeV}
  \\  \cline{3-4} \cline{5-6} \cline{7-8}
  Level & keV &$\sigma$, barn & $\sigma_{\rm eff}$, barn      & $\sigma$, barn & $\sigma_{\rm eff}$, barn &
   $\sigma$, barn & $\sigma_{\rm eff}$, barn \\
  \hline\hline

  $3/2^+_1$ & 0.0076 & 1.389(-4) & 1.008(-3) & 2.314(-4) & 1.679(-3) & 9.020(-4) & 6.448(-3) \\

  $5/2^+_2$ & 29.2 &  2.082(-4) &  7.120(-4) & 3.463(-4) & 1.186(-3) & 1.339(-3) & 4.545(-3) \\

  $7/2^+_1$ & 42.4 &  9.696(-3) &  1.236(-2) & 1.613(-2) &  2.058(-2) & 6.202(-2) & 7.894(-2) \\

  $7/2^+_2$ & 71.8 &  1.151(-4) &  2.977(-4) & 1.916(-4) & 4.966(-4) & 7.316 (-4)& 1.890(-3) \\

  $9/2^+_1$ & 97.1 &  3.370(-3) &  3.371(-3) & 5.620(-3) & 5.621(-3) & 2.138(-2) & 2.139(-2) \\

  $9/2^+_2$ & 125.4 & 2.291(-5) &  2.291(-5) &3.831(-5) & 3.831(-5) & 1.429(-4) & 1.429(-4) \\
  \hline\hline
  \end{tabular}
  \end{table}

For the $E2$ Coulomb excitation we have
\begin{eqnarray} 
&& \hspace*{-0.5cm}\frac{d\sigma_{E2}(\xi,\vartheta)}{d\,\Omega}=\left(\frac{Z_1e^2}{\hbar v}\right)^2
\frac1{a^2}\,B(E2\uparrow) \frac{df_{E2}(\xi,\vartheta)}{d\Omega}\,,
\nonumber \\
&& \hspace*{-0.5cm} a\approx 0.072\,\frac{Z_1Z_2}{E_{1}({\rm MeV})}(1+A_1/A_2)\cdot 10^{-12}{\rm cm} \,, \
\xi=\frac{Z_1Z_2A_1^{1/2}(1+A_1/A_2)\Delta E}{12.65(E_1-1/2\cdot\Delta E)^{3/2}}\,.
\end{eqnarray}

Here, $A_1$, $Z_1$ and $E_1$ refer to the projectile, $E_1$ and
$\Delta E$ are energy in the laboratory system and the excitation energy in MeV, $a$ is half the
distance of the closest drawing in the backward scattering. Functions $f_{E2}(\xi,\vartheta)$
are expressed \cite{Alder561} via  integrals over trajectories.
If $\Delta E/E_{1} = 0$, then  we obtain

\begin{equation} 
\frac{df_{E2}(\xi=0,\vartheta)}{d\,\Omega}\ =\ \frac\pi{25}
\left\{\Big[1-\frac{\pi-\vartheta}2 \tan\frac\vartheta2\Big]^2\cdot \frac1{\cos^4\vartheta/2}
+ \frac13\right\}.
\end{equation}

In a general case, we have \cite{Alder561}

\begin{equation} 
\sigma_{E2}(\xi)\ =\ 4.78\,
\frac{A_1E_{kin}(A_1,\mbox{MeV})B(E2\uparrow,\mbox{barn}^2)}{Z^2_2(1+A_1/A_2)^2} f_{E2}(\xi)\ \mbox{barn},
\end{equation}
where $f_{E2}(\xi=0)$ = 0.895.

For the $M1$ Coulomb excitation we have
\begin{equation} 
\frac{d\sigma_{M1}(\xi,\vartheta)}{d\,\Omega}=\left(
\frac{Z_1e^2}{\hbar c}\right)^2
\frac{\lambda\hspace*{-0.35cm}-_c(p)^2}4\, B(M1\uparrow)
\frac{d f_{M1}(\xi,\vartheta)}{d\,\Omega}\,,
\quad \lambda\hspace*{-0.35cm}-_c(p)=\frac{\hbar}{m_pc}\,.
\end{equation}

 For $\xi = 0$ we obtain
\begin{equation} 
\hspace{0.5cm} \frac{df_{M1}(\xi=0,\vartheta)}{d\,\Omega}\ =\
\frac{16\pi}9\, \frac{[1-(\pi\!-\!\vartheta)/2\,  \cdot\,\tan \vartheta/2]^2}{\sin^2\vartheta}\, .
\end{equation}

We see from Eq.(21), that by $\Delta E \to 0$ (as in our case) and $\vartheta \to$ 0 the total cross section
logarithmically diverges. At the same time,
the probability of the $M1$ excitation by $\xi = 0$, $P(M1,\xi=0, \vartheta)= d\sigma(M1,\vartheta)/
d\sigma(\rm {Coul},\vartheta) \sim \vartheta^2$ by $\vartheta \to 0$. Thus, the divergence of the $M1$
cross section at $\vartheta \to 0$ is due only to the divergence of the Coulomb scattering at $\vartheta \to 0$,
in this case the the colliding nuclei are far from each other, and the  Coulomb interaction between nuclei
is really screened by the electron clouds. Really, almost all electron charge
of atom is located at distances less than the Bohr radius $R_B=\hbar^2/(m_ee^2)$. In this way, we should
exclude intervals more than $R_{\rm max}$, i.e. exclude scattering angles less than
$\vartheta_{\rm min}$, where

\begin{equation} 
\vartheta_{\min}=2\arcsin\Big(\frac1{R_{\max}/a-1}\Big) \approx\, \frac{2a}{R_{\max}}\,,
 \quad R_{\max}\approx2 R_B\,.
\end{equation}

Then, we obtain

\begin{equation} 
\sigma_{M1}=0.589\cdot10^{-8} Z^2_1\, B(M1\uparrow)\ f_{M1}(\xi=0,\vartheta_{\min})\ \mbox {barn}\,.
\end{equation}

Here, $B(M1)$ is in the units of $\mu^2_N$ and

\begin{equation} 
f_{M1}(\xi=0,\vartheta_{\min})\ =\ \frac{32\,\pi^2}9 \int\limits^\pi_{\vartheta_{\min}}
\frac{[1-(\pi\!-\!\vartheta)/2\,\cdot  \tan \vartheta/2]^2}{\sin \vartheta}\, d\vartheta\, .
\end{equation}

  For $\vartheta_{\min 1,2}$ \,\, less than $1^0$  we have
\begin{equation} 
  f_{M1}(\xi=0,\vartheta_{\min1})\, \approx f_{M1}(\xi=0,\vartheta_{\min2})
  + \frac{32\pi^2}9  \ln \bigg( \frac{\vartheta_{\min2}}{\vartheta_{\min1}} \bigg).
\end{equation}

For protons and $\alpha$-particles with energies 10 MeV bombarding $^{229}$Th,
$\vartheta_{min} \sim 0.1^0$ and $f_{M1} (\xi=0, \vartheta_{min} = 0.1^0) = 186$.
The corresponding cross section is negligible as compared to the $E2$
excitation, this statement is even more valid for excitation of high-lying states,
for which the magnitude of $f_{M1}$ rapidly decreases, see also \cite{Alder562}.
Thus, all levels considered by us here, are populated in the Coulomb excitation by
means of the $E2$ transitions.

One should allow for the fact that settlement of the lowest $3/2^{+}$ level may
happen not only due to the direct Coulomb excitation from the ground state, but also due
to the discharging of the excited higher-lying states. This process is very important as
many of these states are actively excited due to large $B(E2)$ values. In this way, we
took into account excitation of all levels shown in Fig.\,1, as well as all possible $E2$
and $M1$ transitions between them. Corresponding $B(E2)$ and $B(M1)$ values were borrowed by
us from Tables 1 and 2, while the necessary conversion coefficients were borrowed from
\cite{bricc}.
Results of our calculations of cross sections are demonstrated in Table 4. Here, $\sigma$
corresponds to the direct excitation, while $\sigma_{\rm eff}$ is the effective cross section,
that includes settlement of the $3/2_1^+$ state by $\gamma$-transitions from the high-lying levels.
One can easily see that the allowance of feeding from the high-lying states leads to
considerable increase of population of the isomeric state. Note, that taking into account
additional excited states leads to further increase of $\sigma_{\rm eff}$ as compared to $\sigma$.

For example, let's take the foil of $^{229}$Th with thickness $d = 10\, \mu m$.
The density $\rho$ of Th is about $3\cdot10\,^{22}$ atoms/cm$^3$.
Suppose that we have constant in time beam of 10 MeV protons with a beam current $j$ equal to
$1 \mu$A ($\sim 0.6\cdot10^{13}$ atoms/s).Then, the counting rate for transitions
from the 0.0076 keV level (allowing also for the settlement of this level from the
high-lying states that are excited in the process of the Coulomb excitation) is
$N=j\cdot \sigma_{\rm eff}\cdot \rho\cdot d$ $\approx 3\cdot 10\,^5\,\, {\rm s}^{-1}$.
However, this level decays mainly by the electron conversion
$(\alpha^{M1}_{tot} \approx 1.4\cdot 10\,^9)$. Thus, the counting rate for $\gamma$-quanta is only
$N_{\gamma} \sim 2\cdot 10^{-4}\,\, {\rm s}^{-1}$, i.e. $ \sim 20\,\,{\rm d}^{-1}$.
However, one should keep in mind that metallic Th is not transparent for ``blue''
$\gamma$-rays. Thus, it is better to use a target from the radiolucent glassy material
containing Th atoms.

\vspace{0.5cm}
The author acknowledge M.B. Trzhaskovskaya for discussions and calculations concerning
problems of atomic structure, as well as Yu.\,N. Novikov and A.V. Popov for useful critical remarks.

\vspace{0.3cm}


\end{document}